\documentclass[11pt,twoside]{article}

\usepackage{asp2004}
\usepackage{epsf}
\usepackage{psfig}
\usepackage{lscape}
\usepackage{graphicx}

\begin{document}
\title{ Mapping Global Star Formation in the Interacting Galaxy Pair ARP 32}   
\author{I. Damjanov, D. Fadda, F. Marleau, P. Appleton, P. Choi, M. Lacy,
L. Storrie-Lombardi, L. Yan}   
\affil{\it Spitzer Science Center, California Institute of Technology, MC 220-6, Pasadena, CA 91125}   

\begin{abstract} 
A multi-wavelength set of photometric data including UV (GALEX),
optical, near-IR, infrared ({\sl Spitzer}) and radio (VLA 20cm) images and
spectroscopic observations are used to map the dust-obscured and
unobscured star formation in the galaxy pair ARP~32.  The system
consists of an actively star-forming galaxy and another one with
depressed star formation. The most active galaxy has disrupted
morphology and different sites of star formation.  Spectroscopic data
show hints of nuclear activity in its core, intense star formation in
limited regions of the galaxy as well as an underlying population of
stars witnessing a past episode of star formation.  Current star
formation rates are estimated from UV and bolometric IR luminosities.
\end{abstract}


\section{Introduction}

\begin{figure}[t!]
\begin{center}
\includegraphics[scale=0.7]{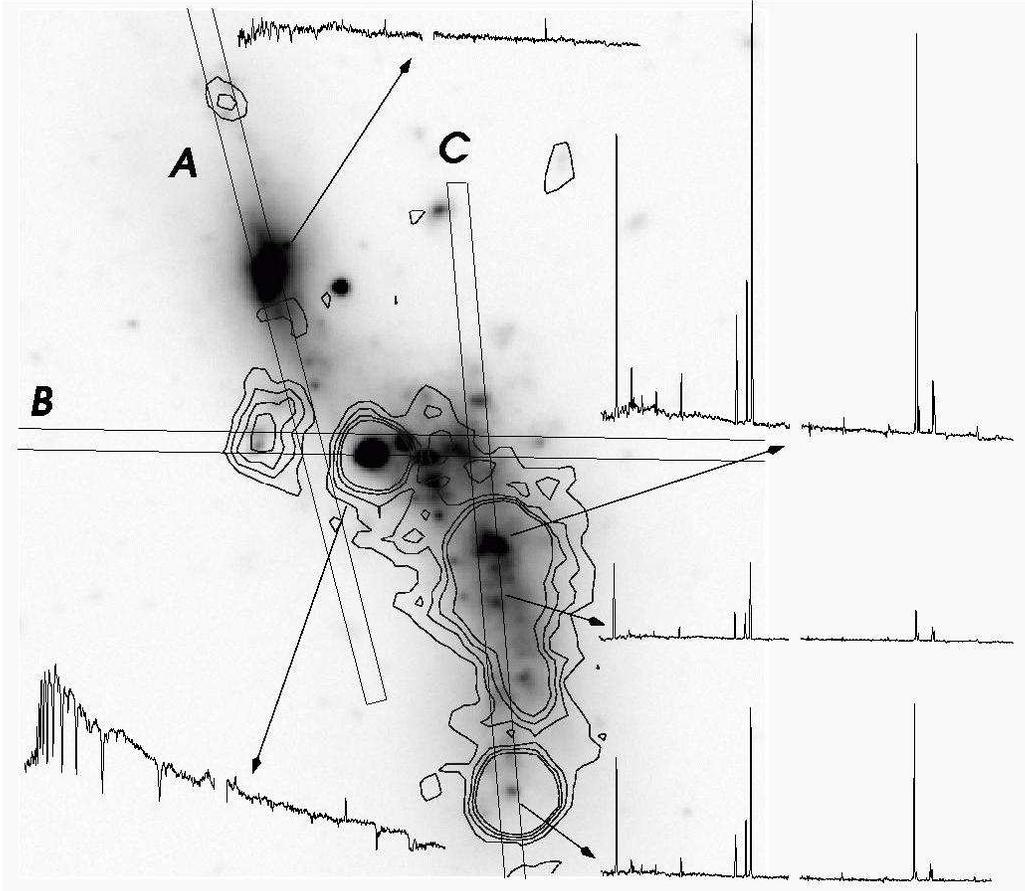}
\end{center}
\caption{ARP~32 R-band image with 24 $\mu$m contours and positions of
the long slits superimposed;  extracted spectra for some of the
emission knots are also presented. The spectrum of the nucleus  of the
upper galaxy shows almost no activity, while the emission lines are
prominent in the  nuclear part of the lower companion. Absorption
features are visible in the central part of the object, in the slit
B.}
\label{fig:fig1}
\end{figure}

The ARP~32 object ($z=0.004$, distance $\approx 17.2$~Mpc) consists of
two interacting galaxies separated by 40$\arcsec$ ($\approx
3.3$~Kpc). The system lies in the field observed with {\sl Spitzer}, VLA,
SDSS and GALEX known as First-Look Survey (see, e.g.,
\citeauthor{Fad2004}\citeyear{Fad2004}). Further ancillary data were
obtained with the Palomar 200'' telescope: J and Ks images with WIRC
and long-slit spectra with Double-spec.  Finally, a {\sl Spitzer}
observation at 16 $\mu$m, taken with the IRS blue peak-up array for
calibration purposes, was used. Our extended photometric dataset (UV
at 1528{\rm \,\AA} and 2271{\rm \,\AA}, u',g',r',i',z', J, Ks, 3.6,
4.5, 5.8 and 8.0 $\mu$m IRAC, 16 $\mu$m IRS, 24, 70 and 160 $\mu$m MIPS
and 20 cm VLA) allows us to map the star formation in the system with
different techniques. A detailed R image (KPNO-4m,
\citeauthor{Fad2004}\citeyear{Fad2004}, Fig.~\ref{fig:fig1})
was used to identify the knots of infrared emission and to program the
spectroscopic observations.  The object was observed at three
different position angles: P.A.~=~14$\deg$ (slit A), P.A.~=~88.5$\deg$
(slit B) and P.A.~=~4$\deg$ (slit C), with a 2$\arcsec$~slit to get
spectra of all the emission knots visible in the R-band image
(Fig.~\ref{fig:fig1}).  We reduced the spectra with the IRAF
packages {\sl twodspec} and {\sl ccdred} and extracted apertures
corresponding to each knot of emission. The spectra cover the range
3000-8000 \AA \ with dispersions 1.07 \AA \ pixel$^{-1}$ and 2.45 \AA
\ pixel$^{-1}$ for the blue and the red arm, respectively.

\section{Spectroscopy}
We measured emission and absorption line fluxes simultaneously using
an IDL code to obtain emission line fluxes corrected for the
underlying stellar absorption. This mainly affects Balmer lines of the
lower companion and, if uncorrected, can lead to big errors in the
evaluation of reddening and star formation densities.  All line fluxes
were measured using multi-Gaussian profile fitting. That enabled us to
deblend the H$\alpha$-$[$N~II$]\lambda\lambda$6548,6563 triplet and
the $[$S~II$]\lambda\lambda$6716,6731 doublet, which are present in
all the extracted spectra. In order to ensure reliable flux
measurements, we considered only absorption and emission lines with
signal-to-noise ratios (SNR) greater than 3.

Measured fluxes of the emission lines were corrected for dust
reddening (i.e., internal extinction) based on the H$\alpha$/H$\beta$
flux ratio \citep{Cal2001}.  When H$\beta$ fluxes could not be
reliably measured, we estimated 3-$\sigma$ upper limits from the
continuum SNR at $\lambda \sim$~4800~{\rm \,\AA} assuming a width 
similar to that of H$\beta$ in other spectra of the same slit.

\begin{figure}[!]
\begin{center}
\includegraphics[height=5.5cm]{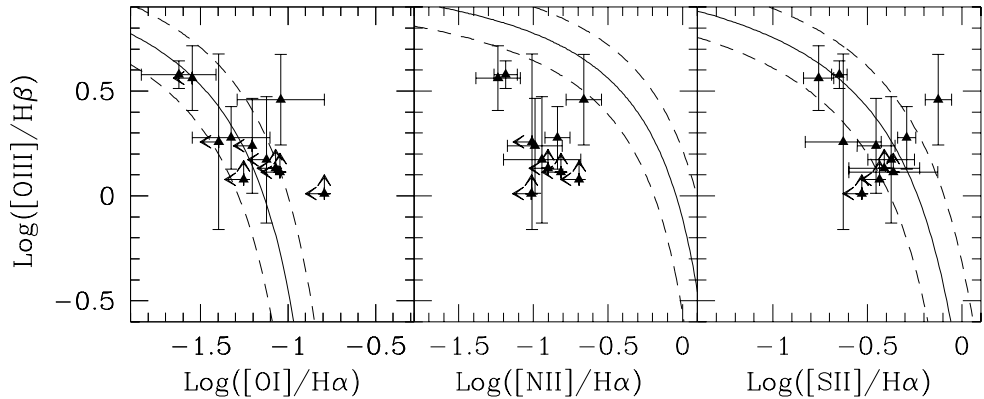}
\end{center}
\caption{VO diagnostics diagrams for the slit C; overplotted curves
were taken from \citet{Kewetal2001}; the
section below the curves in each diagram corresponds to the area
occupied by the H~II-like regions, while the section above the curves
denotes AGN- and LINERs-like features.}
\label{fig:fig2}
\end{figure}

The measured H$\alpha$/H$\beta$ ratio results in negative values for
the broad-band color excess, $E(B-V)$, in some of the infrared
knots. Most of these values are only lower limits since H$\beta$
fluxes could not be reliably measured, but in the nucleus of the lower
companion both lines were detected with high SNR values obtaining a
negative $E(B-V)$ value. Whenever reddening coefficients were
determined to be negative, we assumed the Galactic interstellar
extinction to compute reddening-corrected line flux densities ($E(B-V)
= 0.02$). Corrected line flux densities were used to build
Veilleux-Osterbrock (VO) diagnostic diagrams $[$O~III$]$
$\lambda$5007/H$\beta$ versus $[$O~I$]$ $\lambda$6300/H$\alpha$,
$[$NII$]$ $\lambda$6583/H$\alpha$ and
$[$S~II$]\lambda\lambda$(6716+6731)/H$\alpha$ \Citep*{Kewetal2001}, in
order to study the nature of ionizing sources in emission-line
regions. All emission knots lie in the area of the diagrams occupied
by the H~II-like regions, except for the nucleus of the lower
companion which shows AGN-like features
(Fig.~\ref{fig:fig2}).  Line flux densities were also used to
estimate the metallicity of the emission knots, following the
algorithm given in \Citet{Kewdop2002}. The fact that the nuclear part
of the lower galaxy is metal-poor (log(O/H)$+12< 8.5$) could possibly
explain the negative values of the broad-band color excess $E(B-V)$
(e.g., \citeauthor{Gron2004} \citeyear{Gron2004}).  Deep Balmer
absorption lines are detected in the upper region of the lower galaxy
(see Fig.~\ref{fig:fig1}) which witness a past episode of star
formation.

\section{Photometry}

\begin{figure}[!]
\begin{center}
\includegraphics[scale=0.4]{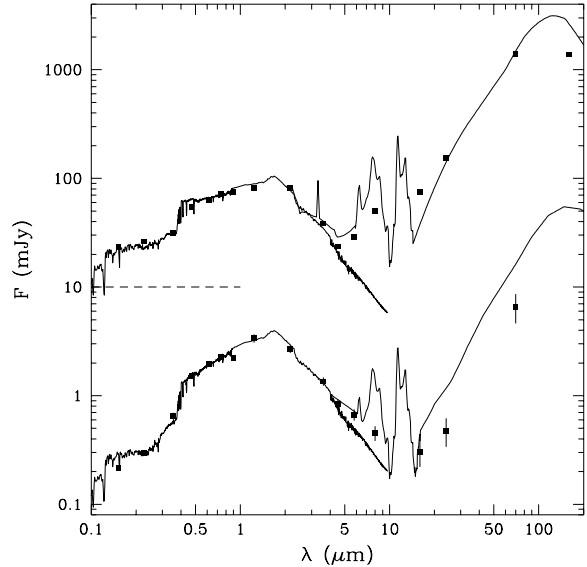}
\end{center}
\caption{SEDs for the two interacting galaxies: the bottom one
corresponds to the upper companion and the top one to the lower
companion. Its flux scale is shifted for presentation purposes and a
dashed line denotes the zero point.}
\label{fig:fig3}
\end{figure}

Aperture photometry was used to measure the fluxes in all the
available passbands for both galaxies.  We corrected the obtained
fluxes for Galactic extinction and use them to build galaxies'
spectral energy distributions (SED) presented in Figure~\ref{fig:fig3}
. In order to compute star formation rates (SFR) based on IR
luminosities, we fitted infrared SEDs with \citet{Chel2001}
templates. The UV, optical and near-IR part were fitted with the
PEGASE~\citep{Frv1997} models. The best fitted model for the upper
galaxy corresponds to an elliptical galaxy with older stellar
population (5 Gyr) and additional 10-30\% of stars younger than 1 Gyr,
while the lower galaxy is best fitted with SO galaxy templates with a
young (0.5 -1 Gyr) stellar population. Following the equations given
in \citet{Ken1998}, we used the measured far-UV and IR fluxes to
estimate the SFRs in both galaxies. The upper companion seems less
active with a SFR value of 0.028~M$_{\sun}$~yr$^{-1}$ according to its
FUV flux and 0.046~M$_{\sun}$~yr$^{-1}$ based on the integrated
8-1000~$\mu$m flux from the \citet{Chel2001} best-fit template. The
latter value is  overestimated since no template fits well the data.
SFR values for the lower galaxy are 0.12~M$_{\sun}$~yr$^{-1}$ and
1~M$_{\sun}$~yr$^{-1}$ based on FUV flux and integrated IR flux,
respectively, showing therefore a large amount of obscured star
formation.
        
\end{document}